\documentclass[prd,preprint,superscriptaddress,amsmath,amssymb]{revtex4}
\usepackage{graphicx,color,slashed}

\usepackage{subfigure,bbold}
\usepackage[toc,page]{appendix}

\def\be{\begin{eqnarray}}
\def\ed{\end{eqnarray}}
\def\non{\nonumber}

\def\ep{\pmb \varepsilon}

\begin{document}

{\begin{flushright}{KIAS-P16019}
\end{flushright}}

\title{ Bounds on LFV Higgs decays in a vector-like lepton model and searching for doubly charged leptons at the LHC}

\author{ Chuan-Hung Chen \footnote{Email: physchen@mail.ncku.edu.tw} }
\affiliation{Department of Physics, National Cheng-Kung University, Tainan 70101, Taiwan }

\author{ Takaaki Nomura \footnote{Email: nomura@kias.re.kr }}
\affiliation{School of Physics, Korea Institute for Advanced Study, Seoul 130-722, Republic of Korea}

\date{\today}

\begin{abstract}
The Higgs-portal lepton flavor violation is studied in a vector-like lepton model. To avoid the constraints from rare $Z\to \ell^\pm_i \ell^\mp_j$ decays, we introduce two triplet vector-like leptons, $(1,3)_{-1}$ and $(1,3)_{0}$. The resultant branching ratio for $h\to \mu \tau$ can be up to $10^{-4}$ when the constraints from the invisible $Z$ decays are applied. As a result,  the signal strength for the $\tau\tau$ channel has a $12\%$ deviation from the standard model prediction, while the muon $g-2$ is two orders of magnitude smaller than the data, and $BR(\tau\to \mu \gamma)$ is of the order of  $10^{-12}$. A predicted doubly charged lepton in $pp$ collisions at $\sqrt{s}=13$ TeV is analyzed, and  it is found that  the interesting production channels are $pp\to (\Psi^{--}_{1} \Psi^{++}_1, \Psi^{\pm\pm}_1 \Psi^\mp_1)$. Both  single and pair production cross sections of $\Psi^{++}_1$ are comparable, and can be   a few hundred fb. The main decay channels for the doubly charged lepton are $\Psi^{\pm\pm} \to \ell^\pm W^\pm$, and for  the heavy singly charged lepton are $\Psi^\pm_1 \to \nu W^\pm, \ell^\pm Z$. The numerical analysis is carried out with regard to $13$ TeV LHC  with $100$ fb$^{-1}$ luminosity. 
\end{abstract}

\maketitle

 Due to a number of  unsolved issues, such as the origin of neutrino mass, dark matter (DM), and matter-antimatter asymmetry, it is believed that the standard model (SM) of particle physics is an effective theory at the electroweak scale. Since rare decays are important in the development of  new physics,  the loop-induced flavor changing neutral current (FCNC) processes are generally used to examine the SM. However, most hadronic processes involve very uncertain non-perturbative quantum chromodynamic (QCD) effects, and thus,  even if new physics exists, it is not easy to distinguish this  from the SM results due to QCD uncertainty.
 
 Leptons do not carry a color charge, and QCD uncertainty is thus much smaller in this case. However, due to the Glashow-Iliopoulos-Maiani (GIM) mechanism,   lepton FCNC processes in the SM (e.g., $\mu\to e \gamma$ and $\tau \to (e,\mu)\gamma$) are highly suppressed; if any signal of lepton flavor violation (LFV) is observed, it is certainly strong evidence for new physics. It is thus  important to search for new physics through the lepton sector~\cite{Crivellin:2013wna,Gomez:2014uha,Crivellin:2015hha}. 

With the discovery  of  the SM Higgs in the ATLAS~\cite{:2012gk} and CMS~\cite{:2012gu} experiments, we are moving   toward better understanding the process of electroweak symmetry breaking (EWSB) through the spontaneous symmetry breaking (SSB) mechanism in the scalar sector. The  next mission for the High Luminosity Large Hadron Collider (LHC) is to explore not only the detailed properties of the SM Higgs, but also the new physics effects.   

 Since the SM Higgs has been discovered, it is of interest to search for new physics  through the Higgs portal. For instance, an excess of events with  a significance of $2.4\sigma$ in $h\to \mu \tau$ decay was reported by CMS in $pp$ collisions at $\sqrt{s}=8$ TeV, where the branching ratio (BR) with the best fit is given by~\cite{Khachatryan:2015kon}:
 \begin{eqnarray}
BR(h\to \mu\tau)=(0.84 ^{+0.39}_{-0.37})\% \  \ [\text{CMS}]\,.
 \end{eqnarray}
ATLAS also reported the same measurement and found no significant excess, where the best fit is~\cite{Aad:2015gha}:
 \begin{eqnarray}
 BR(h\to \mu\tau)=(0.77 \pm 0.62)\% \ \ [\text{ATLAS}]\,.
 \end{eqnarray}
Although the measurements of $BR(h\to \mu \tau)$ are not conclusive yet,  inspired by the Higgs portal events, a number of the possible new physics effects that could explain the large BR for $h\to \mu \tau$ decay have been studied~\cite{Campos:2014zaa,Sierra:2014nqa,Lee:2014rba, Heeck:2014qea,Crivellin:2015mga,Dorsner:2015mja,Omura:2015nja,Crivellin:2015lwa,Das:2015zwa,Bishara:2015cha,Varzielas:2015joa,He:2015rqa,Chiang:2015cba,Altmannshofer:2015esa,Cheung:2015yga,Arganda:2015naa,Botella:2015hoa,Baek:2015mea,Huang:2015vpt,Baek:2015fma,Arganda:2015uca,Aloni:2015wvn,Benbrik:2015evd,Buschmann:2016uzg,Sher:2016rhh,Chang:2016ave,Han:2016bvl}. In this study, we explore the LFV in the framework of a vector-like lepton model.

FCNCs  are quite a common phenomenon in the quark sector, such as neutral meson oscillations and $b\to s \gamma$. However,  with the exceptions of the neutrino oscillations we have no concrete and solid signals to verify the LFV in the lepton sector, thus limiting our knowledge about this. In this context, the measurements from ATLAS and CMS of the Higgs-portal LFV provide a good chance to better understand the lepton sector. 
 Following the hint of the SM with regard to whether the Higgs couplings to the fermions appear in the Yukawa sector, a possible minimal extension of the SM for Higgs mediated LFV is to include exotic heavy leptons or to add a new Higgs doublet without imposing any symmetry~\cite{Benbrik:2015evd}.
In this work,  we study the implications of adding heavy  leptons.
In order to avoid the gauge anomaly, we investigate the model with vector-like leptons (VLLs). 

To achieve mixing with the SM leptons, the representations of VLL under $SU(2)\times U(1)_Y$ gauge symmetry can be singlet, doublet~\cite{Altmannshofer:2013zba,Falkowski:2013jya,Ma:2014zda,Dermisek:2014cia,Dermisek:2014qca,Dermisek:2015vra,Dermisek:2015oja,Kumar:2015tna,Dermisek:2015hue}, and triplet~\cite{Ma:2013tda}. The VLLs from a composite model are discussed in earlier works \cite{Biondini:2012ny,Leonardi:2014epa}. In order to avoid the constraints from rare $Z\to \ell^\pm_i \ell^\mp_j$ decays, we study the triplet representations $(1,3)_{-1}$ and $(1,3)_{0}$ with hypercharges $Y=-1$ and $Y=0$, respectively. 
The new Yukawa couplings are thus written as:
 \begin{eqnarray}
 -{\cal L}_{Y} &= \bar L {\bf Y}_{1} \Psi_{1R} H + \bar L {\bf Y}_{2} \Psi_{2R} \tilde{H} + m_{\Psi_1} Tr\bar \Psi_{1L} \Psi_{1R}  + m_{\Psi_2} Tr\bar \Psi_{2 L} \Psi_{2R} + H.c. \,, \label{eq:Yul}
 \end{eqnarray}
where we have suppressed the flavor indices; $H$ is the SM Higgs doublet, $\tilde H= i\tau_2 H^*$, the neutral component of Higgs field is $H^0 = (v + h)/\sqrt{2}$, and the representations of two VLLs are:
 \begin{eqnarray}
\Psi_{1} = 
\left(
\begin{array}{cc}
 \Psi^-_1/\sqrt{2} & \Psi^0_1    \\
\Psi^{--}_1 &  -\Psi^-_1/\sqrt{2}     
\end{array}
\right)\,, \  \Psi_{2} = \frac{1}{\sqrt{2}}
\left(
\begin{array}{cc}
 \Psi^0_{2}/\sqrt{2} & \Psi^{+}_{2}  \\
\Psi^{-}_{2} &  -\Psi^0_{2}/\sqrt{2}     
\end{array}
\right)
 \end{eqnarray}
with $\Psi^+_{2} = C\bar \Psi^{-}_{2}$ and 
 $ \Psi_{2}^{0} = C \bar \Psi_{2}^{0}$.  Since $\Psi_2$ is a real representation of $SU(2)$, the factor of $1/\sqrt{2}$ in $\Psi_{2}$ is to obtain the correct mass term for Majorana fermion $\Psi^0_2$. 
 Due to the new Yukawa terms of ${\bf Y}_{1,2}$, the heavy neutral and charged leptons mix with the SM leptons; after EWSB, the lepton mass matrices become $5\times 5$ matrices and are expressed by:
 \begin{eqnarray}
 M_{\ell} = \left(
\begin{array}{cc}
 {\bf m}_{\ell}  & {\bf Y}^\ell v    \\
0 & {\bf m}_{\Psi}     
\end{array}
\right)\,, \  M_{\nu} = \left(
\begin{array}{cc}
 {\bf m}_{\nu}  & {\bf Y}^\nu v    \\
0 & {\bf m}_{\Psi}     
\end{array}
\right)\,, \label{eq:masses}
\end{eqnarray}
where we have chosen the basis such that  the SM leptons are in diagonalized states, ${\bf m}_{\ell}$ is the SM charged lepton mass matrix, ${\bf m}_{ \Psi}=$diag$(m_{\Psi_1}, m_{\Psi_2})$, and 
\begin{eqnarray}
 {\bf Y}^\ell = \frac{1}{2} \left(
\begin{array}{cc}
 -Y_{11} &  Y_{21}    \\
- Y_{12} &   Y_{22} \\
 -Y_{13} &  Y_{23}     
\end{array}
\right)\,, \  {\bf Y}^\nu= \sqrt{2} \left(
\begin{array}{cc}
 Y_{11} & Y_{21}/2   \\
 Y_{12} &  Y_{22}/2 \\
Y_{13}& Y_{23}/2     
\end{array}
\right)\,. \label{eq:Ylnu}
 \end{eqnarray}
 We note that the elements of ${\bf Y}^\chi$ should be read as $Y_{ij}= ({\bf Y}_i)_j$, where the index $i=1,2$ distinguishes the Yukawa couplings of the different VLLs and the index $j=1,2,3$ stands for the  flavors of the SM leptons. 
 Since the origin of neutrino mass has not been concluded yet and is still model dependent, we directly put the Majorana type of neutrino mass term to the Yukawa sector. Since the detailed effects of neutrino physics are irrelevant to this study, we do not further pursue  issues related to this  and ${\bf m}_\nu = 0$.  

To diagonalize $M_{\ell}$ and $M_{\nu}$, we  introduce the unitary matrices $V^{\chi}_{R,L}$ with $\chi=\ell, \nu$ so that $M^{\rm dia}_\chi= V^\chi_L M_{\chi} V^{\chi \dagger}_R$. The information of $V^\chi_L$ and $V^\chi_R$ can be obtained through $M_\chi M^\dagger_\chi$ and $M^\dagger_\chi M_\chi$, respectively.  According to Eq.~(\ref{eq:masses}), it can be easily found  that the flavor mixings  between heavy and light leptons in $V^\chi_{R}$ are proportional to the lepton masses. Since the neutrino masses are small, it is a good approximation to take $V^\nu_R \approx 1$. If we further set $m_e = m_\mu =0$ in the phenomenological analysis, only $\tau$-related processes have significant contributions. Unlike $V^\chi_R$, the off-diagonal elements in flavor-mixing matrices $V^\chi_L$ are associated with ${\bf Y}_{1,2} v/{\bf m}_\Psi$; in principle, the mixing effects can be of the order of $0.1$. In this study, we examine these effects on $h\to \tau \mu$ and rare tau related decays. To be more specific, we parametrize the unitary matrices in terms of ${\bf Y}_{1,2}$ as:
 \begin{align}
 V^\chi_{L} \approx   \left(
\begin{array}{cc}
 \mathbb{1}_{3\times 3} - \ep^\chi_L \ep^{\chi\dagger}_L / 2  &  - \ep^\chi_L    \\
\ep^{\chi\dagger}_L &  \mathbb{1}_{3\times 3} - \ep^{\chi \dagger}_L \ep^{\chi}_L/2  
\end{array}
\right)\,,\ V^\ell_{R} \approx   \left(
\begin{array}{cc}
 \mathbb{1}_{3\times 3} &  - \ep^\ell_R    \\
\ep^{\ell\dagger}_R &  \mathbb{1}_{3\times 3}  
\end{array}
\right)\,, \label{eq:VRL}
 \end{align}
where  $V^\nu_R  \approx 1$ is implied, $\ep^\chi_L \approx v {\bf Y}^\chi/{\bf m}_\Psi$, and $\ep^\ell_R \approx v {\bf m}^\dagger_{\ell} {\bf Y}^\ell/{\bf m}^2_{\Psi}$.

Combining the SM Higgs couplings and new Yukawa couplings of Eq.~(\ref{eq:Yul}), 
the Higgs couplings to all singly charged leptons are given by:
 \begin{align}
 -{\cal L}_{h\ell' \ell'} &= h \bar \ell'_L V^\ell_L \left(
\begin{array}{cc}
 {\bf m}_\ell /v  & Y^\ell    \\
0 & 0    
\end{array}
\right) V^{\ell \dagger}_{R} \ell'_R+ H.c.\,, 
 \end{align}
where  $\ell'^T = (e, \mu, \tau, \tau', \tau'')$ is the state of a physical charged lepton in lepton flavor space. We use the notations of $\tau'$ and $\tau''$ to denote  the  heavy charged  VLLs. Using the parametrization of Eq.~(\ref{eq:VRL}), the Higgs couplings to the SM leptons can be formulated by:
 \begin{align}
 -{\cal L}_{h \ell \ell} & = C^{h}_{ij} \bar\ell_{iL} \ell_{jR} h + H.c.\,, \label{eq:hll}\\ 
 C^{h}_{ij} & = \frac{m_{\ell j}}{v} \left[ \delta_{ij} - \frac{3}{8} \left( \frac{v^2 Y_{1i} Y_{1j}}{m^2_{\Psi_1}} +\frac{v^2 Y_{2i} Y_{2j}}{m^2_{\Psi_2}}\right) \right]\,. \non
 \end{align}
If we set $m_e=m_\mu=0$, it is clear that in addition to  the coupling $h\tau\tau$ being modified, the tree-level  flavor-changing couplings $h$-$\tau$-$\mu$ and $h$-$\tau$-$e$ are induced, and the couplings are proportional to $m_\tau/v\approx 7.2\times 10^{-3}$. In order to study the VLL contributions to $h\to \gamma\gamma$,  the couplings for $h\tau'\tau'$ and $h\tau'' \tau''$ are expressed as:
 \begin{align}
 -{\cal L}_{h\Psi\Psi} &= \frac{v \sum_{i} Y^2_{1i}}{2 m_{\Psi_1}}  h \tau' \tau' +  \frac{v \sum_{i} Y^2_{2i} }{2 m_{\Psi_2}}  h \tau'' \tau'' \,. \label{eq:hLL}
 \end{align}

Due to the mixture between VLLs and the SM leptons, the same effects also influence the gauge couplings of $Z$ and $W$ to the SM leptons. To understand the modifications, we discuss the gauge interactions below. With the covariant derivative for triplet VLLs \cite{Chen:2015cfa}, we first write the $Z$-boson gauge interactions with VLLs to be:
 \begin{align}
 {\cal L}_Z & =  -\frac{g}{c_W} Z_\mu \left[ \bar \Psi_L \gamma^\mu \left( I^\Psi_3 -s^2_W Q_\Psi \right) \Psi_L + \bar \Psi_R \gamma^\mu ( -s^2_W Q_\Psi ) \Psi_R + \overline{\Psi^{--}_1} \gamma^\mu (-1 + 2 s^2_W) \Psi^{\rm --}_1  \right. \non \\
 & \left. +  \bar \Psi_L \gamma\mu  \left(\begin{array}{cc}
1/2  &  0  \\
0 & -1/2 
\end{array}
\right) \Psi_L
+ \bar N_R \gamma^\mu  \left(\begin{array}{cc}
1/2  &  0  \\
0 & 0  
\end{array}
\right) N_R  + \bar L'_R \gamma^\mu  \left(\begin{array}{cc}
0  &  0  \\
0 & -1
\end{array}
\right) L'_R \right] \,, \label{eq:intZ}
 \end{align} 
where we have expressed the forms of couplings to be the same as  those  in the SM, $I^\Psi_3 =1/2 (-1/2)$ and $Q_\Psi=0 (-1)$ for neutral (charged) VLLs, $\Psi^T=(\Psi^0_1, \Psi^0_2)$ or $(\Psi^-_1 , \Psi^-_2)$, $N^T_R = (\Psi^0_{1R}, \Psi^0_{2R})$, and $L^{'T}_R = (\Psi^-_{1R}, \Psi^{-}_{2R})$. It is clear that the first two terms in Eq.~(\ref{eq:intZ}) provide the flavor-conserving effects; however, the last three terms lead to $Z$-mediated FCNC couplings at the tree level. As mentioned earlier, the flavor mixings in $V^\chi_R$ are associated with the lepton masses; if we ignore the small effects, the gauge couplings of the $Z$-boson to the neutral and singly charged leptons can be expressed as:
\begin{align}
{\cal L}_{Z\ell' \ell'} &= -\frac{g}{c_W} C^{\ell'_L}_{ij} \bar\ell'_L \gamma^\mu \ell'_L Z_\mu - \frac{g}{c_W} C^{\ell'_R}_{ij} \bar\ell'_L \gamma^\mu \ell'_R Z_\mu \,, \label{eq:ZFCNC}\\
C^{\ell'_L}_{ij} & =( I^{\ell'}_3 -s^2_W Q_{\ell'}) \delta_{ij} + \frac{1}{2} \left( V^{\ell'}_{Li4} V^{\ell' *}_{Lj4} - V^{\ell'}_{Li5} V^{\ell' *}_{Lj5} \right) \,,  \non \\
C^{\ell'_R}_{ij}  & \approx -s^2_W Q_{\ell'} \delta_{ij} + \left\{\begin{array}{c} 
     \frac{1}{2}\delta_{i4} \delta_{j4} \ \text{for } Q_{\ell'}=0 \\
     - \delta_{i5} \delta_{j5}  \ \text{for } Q_{\ell'}=-1 
     \end{array} \right.
     \non 
 \end{align}
with $\ell'^T  = ( \nu _1, \nu_2, \nu_3, \nu_4, \nu_5)$ or $(e, \mu , \tau, \tau', \tau'')$. As a result, the tree-level $Z$-mediated FCNCs only occur in the left-handed currents. 

The new gauge interactions of the $W$-boson with VLLs are given by:
 \begin{eqnarray}
 {\cal L}_W = & - g \left( \overline{ \Psi^0_1} \gamma^\mu \Psi^-_1 + \overline{ \Psi^-_1 } \gamma^\mu \Psi^{--}_1 \right) W^+_\mu  - g \left( \overline{\Psi^0_2} \gamma^\mu \Psi^-_2 \right) W^+_\mu + H.c.\,. \label{eq:IntW}
 \end{eqnarray}
With $V^\chi_R \approx 1$, the $W$-mediated interactions of neutral and singly charged leptons are expressed as:
 \begin{eqnarray}
 {\cal L}_{W\nu' \ell'} = - \frac{g}{\sqrt{2}} \overline{{\cal N}_L} \gamma^\mu V^\nu_L  
 \left(\begin{array}{cc}
V'_{\rm PMNS} &  0  \\
0 &  \sqrt{2}
\end{array}
\right)  V^{\ell \dagger}_L  \ell'_L W^+_\mu + H.c. \,, \label{eq:W}
 \end{eqnarray}
where $V'_{\rm PMNS}$ is the Pontecorvo-Maki-Nakagawa-Sakata (PMNS) matrix without triplet VLLs, which can be regarded as a model-dependent result, and is uncertain. Since ${\bf m}_\nu=0$, in this study $V'_{\rm PMNS}=1$. With the flavor mixings in Eq.~(\ref{eq:VRL}), from Eq.~(\ref{eq:W}) the $W$-boson interactions with the SM leptons are formulated by:
 \begin{align}
 {\cal L}_{W\nu \ell}  & = -\frac{g}{\sqrt{2}} ( \bar \nu_1, \bar \nu_2, \bar \nu_3)_L \gamma^\mu V_{\rm PMNS}  \left(\begin{array}{c}
   e \\
   \mu \\
   \tau \end{array} \right)_L W^+_\mu + H.c.\,, \\
   V_{\rm PMNS} &= V'_{\rm PMNS} - V'_{\rm PMNS} \frac{{\ep}^\ell_L {\ep}^{\ell \dagger}_{L}}{2} - \frac{{\ep}^\nu_L \ep^{\nu \dagger}_L}{2} V'_{\rm PMNS}  + \sqrt{2} {\ep}^\nu_L {\ep}^{\ell \dagger}_L\,. 
 \end{align}
The $V_{\rm PMNS}$ is the $3\times 3$ PMNS matrix, which can be extracted from the matrix product of $V^\nu_L (...) V^{\ell \dagger}_L$ in Eq.~(\ref{eq:W}). 
Since the  minimal value of the PMNS matrix element is around $0.15$~\cite{PDG}, 
 the limits of $Y_{1i}$ and $Y_{2i}$ from the charged current interactions may not be as clear as those from the rare $Z$ decays. Therefore, to constrain the free parameters, we focus on the rare $Z$ decays, such as $Z\to \ell^{\pm}_i \ell^{\mp}_j$, invisible $Z$ decays, and so on.

\begin{figure}[hptb] 
\begin{center}
\includegraphics[width=75 mm]{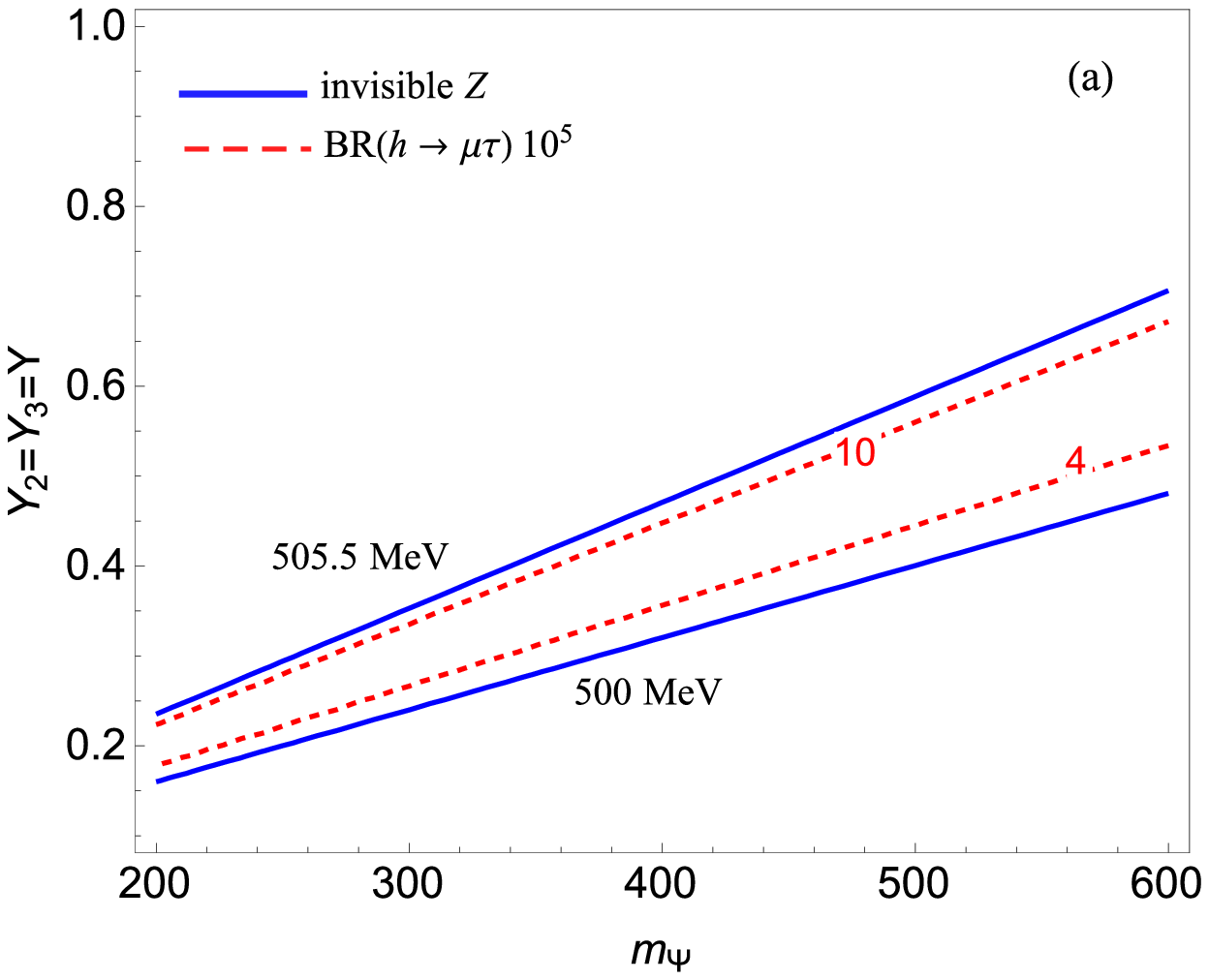} 
\includegraphics[width=75 mm]{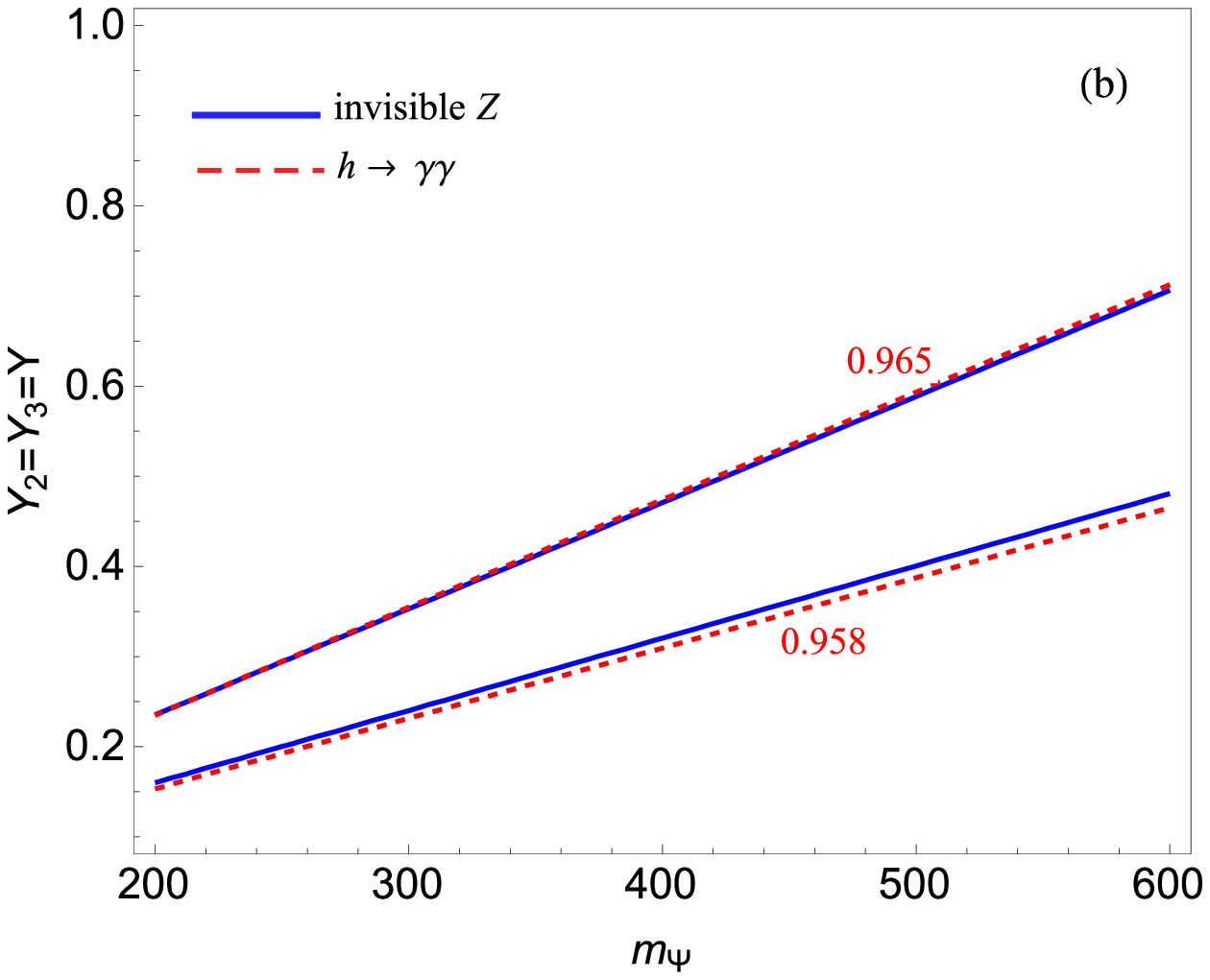} 
\caption{ Contours as a function of $Y$ and $m_{\Psi}$: (a)  for $\Gamma(Z\to \nu_i \nu_j)$ and $BR(h\to \mu\tau)$ and (b) for  signal strength $\mu_{\gamma\gamma}$, where the numbers on the plot (a) for $BR(h\to \mu\tau)$ are in units of $10^{-5}$.}
\label{fig:invZ}
\end{center}
\end{figure}

Before studying the LFV-related phenomenologies, we discuss the possible constraints on the free parameters $Y_{1i,2i}$. From Eq.~(\ref{eq:ZFCNC}), it can be seen that the $Z$-mediated lepton flavor-violating effects can contribute to $Z\to (e \mu, e\tau, \mu \tau)$, where the current upper limits of the data are~\cite{PDG}:
  \begin{align}
 & Br(Z\to e^\pm \mu^\mp) < 7.5\times 10^{-7}\,, \non \\
  & Br(Z\to e^\pm \tau^\mp) < 9.8 \times 10^{-6}\,, \non \\
  & Br(Z\to \mu^\pm \tau^\mp ) < 1.2 \times 10^{-5} \,.
  \end{align}
The severe constraints make the BR of $h\to \tau \mu$ decay far smaller than the CMS measurements. In order to avoid the bounds from the rare $Z$ decays, we set $Y_{1i} =  Y_{2i} $ and $m_{\Psi_1} = m_{\Psi_2} $. As a result, the second term of $C^{\ell'_L}_{ij}$  in Eq.~(\ref{eq:ZFCNC})  for charged leptons vanishes. However, the cancellations are not complete in the $Z\to \bar \nu_i \nu_j$ decays, due to the structure of ${\bf Y}^\nu$ in Eq.~(\ref{eq:Ylnu}). That is, the invisible $Z$-boson decays can directly constrain the parameters, where the current measurement is $\Gamma^Z_{\rm inv}=499 \pm 1.5 $ MeV~\cite{PDG} and the SM prediction is around 500 MeV. With $Y_{1i} =  Y_{2i} = Y_i$ and $m_{\Psi_1} = m_{\Psi_2} = m_{\Psi}$, the partial decay rates for $Z\to \bar\nu_i \nu_j$ and $h\to \mu \tau$ are given as:
 \begin{align}
 \Gamma(Z\to \bar \nu_i \nu_j) & \approx  \frac{m_Z}{24 \pi} \frac{g^2}{c^2_W} \sum_{ij} |C^\nu_{ij}|^2\,, \\
 %
 %
  \Gamma(h\to \mu \tau) & \approx \frac{m_h}{16\pi } \frac{m^2_\tau}{v^2} \left| \frac{ 3v^2 Y_2 Y_3}{2 m^2_\Psi}\right|^2\,.
 \end{align}
Accordingly, we present the contours for $\Gamma(Z\to \bar\nu_i \nu_j)$ and $BR(h\to \tau \mu)$ as a function of $Y_i$ and $m_\Psi$ in Fig.~\ref{fig:invZ}(a), where numerically we adopt $Y_2=Y_3=Y$ and $Y_1\ll 1$,  the solid  line stands for $Z\to \bar \nu_i \nu_j$, the dashed line is for $h\to \mu \tau$ and the values on the plot are in units of $10^{-5}$, and $\Gamma_h \approx 4.21$ MeV is used. The results clearly show that the lepton flavor-violating Higgs decay can only be up to $10^{-4}$ when the data for invisible $Z$ decays are applied.

 Next, we discuss the influence of new flavor mixings on other Higgs decays.  From Eq.~(\ref{eq:hLL}), it can be seen that the induced couplings of Higgs to VLLs can contribute to $h\to \gamma\gamma$ through the loop diagrams; the decay rate can be formulated as:
   \begin{align}
\Gamma(h\to \gamma\gamma) & \approx  \Gamma^{\rm SM}(h\to \gamma\gamma) \left| 1 + \frac{C_{VLL}}{A_W + N_C Q^2_t A_{t}} \right|^2\,, \\
%
C_{LLV} & =  \frac{v^2\sum_i Y^2_{i} }{2 m^2_{\Psi}} A_{1/2}(\tau_{\Psi}) 
\end{align}
where $N_C =3$, $Q_t = 2/3$, $A_W \approx 8.3 $, $A_t \approx -1.38$, and the loop integral from VLL is~\cite{Gunion:1989we}:
\begin{equation}
A_{1/2}(\tau) =  -2 \tau [1+(1-\tau) f(\tau)^2] \non 
  \end{equation}
  with $\tau_\Psi= 4 m_{\Psi}^2/m_h^2$ and $f(x)=\sin^{-1}(1/\sqrt{x})$.  We note that although the coupling $h\tau\tau$ is modified by the new flavor mixing effects, we ignore its small contribution to the loop-induced $h\to \gamma\gamma$ decay. The signal strength, which is used to show the Higgs measurement, is defined as:
   \be
   \mu_{\gamma\gamma} = \frac{\sigma(pp\to h)}{\sigma(pp\to h)_{\rm SM}} \times \frac{BR(h\to f)}{BR(h\to f)_{\rm SM}}\,,
   \ed
where $f$ denotes the possible decay channel. Taking $\Gamma_h \approx 4.21$ MeV and $\sigma(pp\to h) = \sigma(pp\to h)_{\rm SM}$, we plot the contours for $\mu_{\gamma\gamma}$ as a function of $Y$ and $m_\Psi$ in Fig.~\ref{fig:invZ}(b). For comparison, we also show the constraint from $\Gamma^Z_{\rm inv}$ in the same plot. In these results we see that the deviation from the SM prediction is about $4\%$ and is consistent with   $\mu_{\gamma\gamma}=1.17\pm 0.27$ and $1.13\pm 0.24$, as measured by ATLAS~\cite{Aad:2015gba}   and CMS~\cite{CMS:2014ega}, respectively.
 
 From Eq.~(\ref{eq:hll}),  it can be seen that the modified Higgs couplings to the SM leptons are  still proportional lepton masses. By comparison with other lepton channels, it can be seen that the $\tau\tau$ mode is more significant, and thus   we study  the influence on $h\to \tau^+ \tau^-$. Using the values that satisfy $BR(h\to \mu \tau) \approx 10^{-4}$,  the deviation of $\Gamma(h\to \tau^+ \tau^-)$ from the SM results can be obtained as:
 \begin{eqnarray}
 \kappa_{\tau\tau} \equiv \frac{\Gamma (h\to \tau^+ \tau^-)}{\Gamma^{\rm SM} (h\to \tau^+ \tau^-)} = \left|1- \frac{6 v^2 Y^2_3}{8 m^2_\Psi} \right|^2 \approx 0.88\,.
 \end{eqnarray}
 If the SM Higgs production cross section is not changed, the signal strength for $pp\to h\to \tau^+ \tau^-$ in this model is $\mu_{\tau\tau} \approx 0.88$, where the measurements from ATLAS and CMS are $1.44^{+0.42}_{-0.37}$~\cite{Aad:2015gba} and $0.91 \pm 0.27$~\cite{CMS:2014ega}, respectively. Although the current data errors for the $\tau\tau$ channel are still large, the precision measurement of $\mu_{\tau\tau}$ can test the model or give  strict limits on the parameters.
  
\begin{figure}[hptb] 
\begin{center}
\includegraphics[width=4 in]{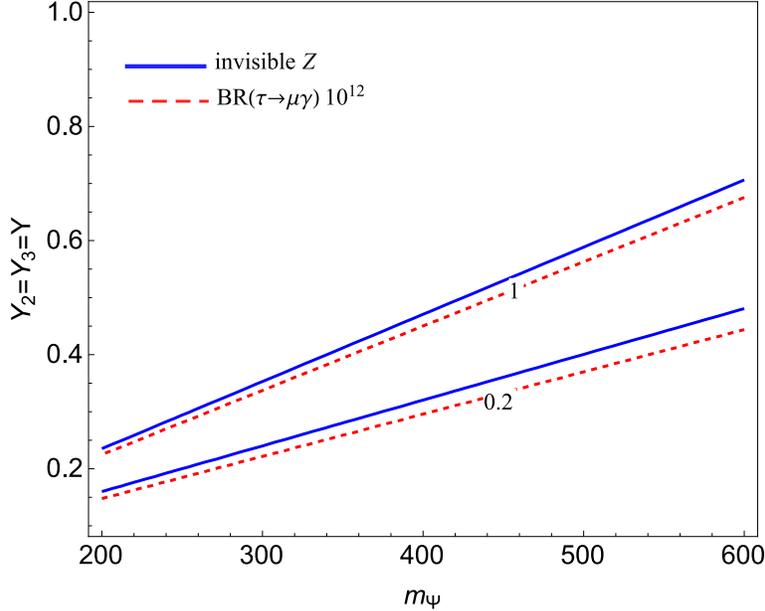} 
\caption{ Contours for $BR(\tau \to \mu \gamma)$ (dashed) as a function of $Y$ and $m_\Psi$, where the constraint from $\Gamma^Z_{\rm inv}$ (solid) is included.   }
\label{fig:taumuga}
\end{center}
\end{figure}

 In the following text we investigate the contributions of new couplings in Eq.~(\ref{eq:hll}) to the rare tau decays and to the flavor-conserving muon anomalous magnetic moment. We first examine the muon $g-2$, denoted by $\Delta a_\mu$. The lepton flavor-changing coupling $h\mu\tau$ can contribute to the $\Delta a_\mu$ through the Higgs-mediated loop diagrams. However, as shown in Eq.~(\ref{eq:hll}), the induced couplings are associated with  $m_{\ell j}/v \bar \ell_{Li} \ell_{Rj}$; that is, only the right-handed tau-lepton has a significant contribution. The induced $\Delta a_\mu $ is thus  suppressed by $m^2_\mu m_\tau /(v m^2_h)$ so that the value of $\Delta a_\mu$ is two orders of magnitude smaller than current data $\Delta a_\mu = a^{\rm exp}_\mu - a^{\rm SM}_\mu =(28.8\pm 8.0)\times 10^{-10}$~\cite{PDG}. A similar situation happens in $\tau\to 3\mu$ decay. Since the couplings are suppressed by $m_\tau/v$ and $m_\mu/v$, the BR for $\tau\to 3\mu$ is of  the order of $10^{-14}$.  We also examine the process $\tau\to \mu \gamma$ via the $h$-mediation. The effective interaction for $\tau \to \mu \gamma$ is expressed by
\be
{\cal L}_{\tau \to \mu \gamma} =\frac{ e }{16\pi^2} m_\tau \bar \mu \sigma_{\mu \nu} \left( C_L P_L + C_R P_R\right) \tau  F^{\mu \nu}\,,
\ed
where $C_L=0$ and  the Wilson coefficient  $C_R$ from the one-loop   is formulated  as:
\begin{align}
C_R & \approx  \frac{ C^h_{23} C^h_{33} }{2m^2_h } \left( \ln \frac{m^2_{h} }{m^2_\tau} - \frac{4}{3}\right) \,. \ \label{eq:WCps}
\end{align}
Accordingly, the BR for $\tau\to \mu \gamma$ is expressed as:
 \begin{eqnarray}
 \frac{BR(\tau \to \mu \gamma)}{BR(\tau \to e \bar \nu_e \nu_\tau) } = \frac{3 \alpha_e}{4\pi G^2_F}  |C_R|^2\,.
 \end{eqnarray}
We present the contours for $BR(\tau\to \mu\gamma)$ as a function of $Y$ and $m_\Psi$ in Fig.~\ref{fig:taumuga}, where the numbers on the plots are in units of $10^{-12}$. It can be seen that the resultant $BR(\tau\to \mu \gamma)$ can be only up to $10^{-12}$, where the current experimental upper bound is $BR(\tau \to \mu \gamma)< 4.4 \times 10^{-8}$~\cite{PDG}.

\begin{figure}[hptb] 
\begin{center}
\includegraphics[width=4 in]{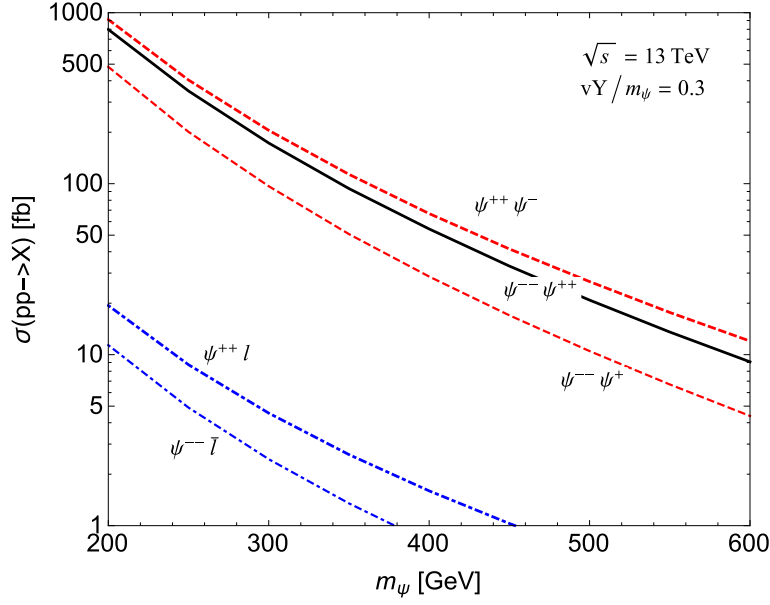} 
\caption{ Doubly charged lepton production cross section ( in units of fb) as a function of $m_\Psi$ in $pp$ collisions at $\sqrt{s}=13$ TeV, where $X$ in the y-axis denotes the possible channel.   }
\label{fig:sigma}
\end{center}
\end{figure}

 In this model, we have two new neutral leptons, two new singly charged leptons, and one doubly charged lepton. Since a particle carrying an electrical charge of 2 can have less background and a clearer signature in colliders, we discuss the potential for discovering  the doubly charged lepton  $\Psi^{--}_1$. By electroweak interactions, $\Psi^{--}_{1}$ can be produced singly and in pairs through the channels $\Psi^{\mp\mp}_1 \Psi^{\pm}_1$ and $\Psi^{--}\Psi^{++}$, where the former is from charged weak  interactions while the latter is from $Z$ and electromagnetic interactions. In addition, due to the flavor mixing effects,  the $W$ couplings to $\Psi^{--}_1$ and the SM leptons can be written as:
\begin{eqnarray}
  {\cal L}_{W\Psi^{--}_{1} \ell} & = - g \bar \ell_i  \gamma^\mu (\ep^{\ell}_{L i 4} P_L + \ep^{\ell}_{R i4} P_R )\Psi^{--}_1 W^+ + H.c.\,. \label{eq:WL2l}
  \end{eqnarray}
By the induced gauge couplings, the doubly charged lepton can be produced through the $\Psi^{--}_{1} \ell$ channels. We note that  due to $\ep^{\ell}_{R} \propto {\bf m}_{\ell}$, the right-handed current contributions can be neglected. In order to discuss the production cross section in $pp$ collisions, we implement our model in CalcHEP~\cite{Belyaev:2012qa} and use   {\tt CTEQ6L} PDF~\cite{Nadolsky:2008zw} to do the numerical calculations. In Fig.~\ref{fig:sigma},  we show the single and pair production cross sections of $\Psi^{\pm \pm}_1$ as a function of $m_\Psi$ in $pp$ collisions at $\sqrt{s}=13$ TeV, where  $vY/m_{\Psi} = 0.3$ is used for the $\Psi^{++}\ell$ production.   It can be seen that the production cross sections for $\Psi^{\pm\pm} \ell$ modes are one order of magnitude smaller than those for other modes. For $m_\Psi < 400$ GeV, the production cross sections for $\Psi^{--} \Psi^{++}$ and $\Psi^{++} \Psi^{-}$, which only  depend on gauge couplings, can be over 50 fb.

Next, we discuss the decays of $\Psi^{--}_1$ and $\Psi^{-}_1$. From Eq.~(\ref{eq:Yul}), we see that before EWSB, the triplet VLLs in  $\Psi_1$ are degenerate; however, the masses are split when the ${\bf Y}_{1}$ effects are involved. Since $\Psi^{--}_{1}$ can not mix with other leptons, the mass splittings occur in neutral and singly charged leptons. According to Eqs.~(\ref{eq:masses}) and (\ref{eq:VRL}), the mass of $\Psi^-_1$ shifted from $\Psi^{--}_1$ is:
 \begin{eqnarray}
m_{\Psi^{-}_1}  \approx  m_{\Psi_1} \left( 1 +  \frac{ v^2 \sum_i Y^2_{1i} }{8 m^2_{\Psi_1}} \right).
 \end{eqnarray}
 With the bound from $\Gamma^Z_{\rm inv}$, the mass splitting is only $3\%$. Therefore, we still take $m_{\Psi^-_1} \approx m_{ \Psi^{--}_1} \approx \Psi^0_1$ in this study. 
 From Eq.~(\ref{eq:WL2l}), we see that the $\Psi^{--}_{1}$ decay pattern depends on each value of Yukawa coupling $Y_{1i}$. If we adopt $Y_{11}\approx Y_{12}\approx Y_{13}=Y$, the better channels to  search for the doubly charged lepton are $\Psi^{--}_1 \to  (e,\mu,\tau) W^-$  and the corresponding  BRs  are fixed as:
   \begin{eqnarray}
   BR(\Psi^{--}_1 \to (e,\mu,\tau) W^-) \approx   1/3\,,
   \end{eqnarray}
   where the lepton masses are ignored. Since $\tau$ has hadronic and leptonic decays and accompanies the neutrino when it decays, the clear signal for probing the doubly charged lepton should be $\Psi^{--}_1 \to \ell W^- \to \ell(\ell \nu)$, where $\ell=e,\mu$ and the final states are the same-sign dilepton. The Higgs and gauge boson couplings of $\Psi^{-}_{1}$ to the SM leptons are given by:
   \begin{eqnarray}
   {I}_{\Psi^-_1} =  \frac{v  Y_{1i}}{2}  \bar \ell_{iL} \Psi^-_{1R}  h + \frac{g}{c_W} \frac{v Y_{1i}}{2m_\Psi}  \bar \ell_{iL} \gamma^\mu \Psi^{-}_{1L} Z_\mu + g \frac{3v Y_{1i} }{2m_\Psi}   \bar \nu_{i L} \gamma^\mu \Psi^{-}_{1L} W^+_\mu +H.c.
   \end{eqnarray}
It is found  that the BRs for $\Psi^-_1 \to (\ell_i h, \ell_i Z, \nu W)$ are insensitive to $Y=Y_{1i}$ and $m_{\Psi}$, and their ratios are :
 \begin{eqnarray}
 \ell_i h : \ell_i Z : \nu W \approx 0.02/3 : 0.1/3 : 0.89\,,
  \end{eqnarray}
  where $\ell_i$ denotes one of the SM leptons and the $\nu W$ channel includes all SM neutrinos. It is clear that the BR for $\nu W$ is one order of magnitude larger than other decay modes. 
  
  According to the analysis, there are two ways to  search for the doubly charged lepton in the model. 
  In pair production, the search channel is
   \begin{eqnarray}
   \label{eq:pair}
   pp\to \Psi^{--}_1 \Psi^{++}_1 \to (\ell^- W^-) (\ell^+ W^+)\,,
   \end{eqnarray}
where $\ell=e,\mu$, the $W$-boson can decay to leptons or jets, and their corresponding cross section with $m_\Psi =300$ GeV is $76$ fb at $\sqrt{s}=13$ TeV.  
The expected events with a luminosity of 100 fb$^{-1}$ are shown in Table~\ref{tab:event}.
In single production, the search channels are  
 \begin{eqnarray}
 \label{eq:single}
 pp\to \Psi^{\pm \pm}_1 \Psi^{\mp}_1 \to \left\{ \begin{array}{c}
                                  (\ell^\pm W^\pm) ( \nu W^\mp )\,, \\
                                  (\ell^\pm W^\pm) ( \ell^\mp Z ) \,,
                                   \end{array}\right. 
 \end{eqnarray} 
 where the associated cross sections for $m_\Psi=300$ GeV at $\sqrt{s}=13$ TeV are $\sigma(\ell^+ \nu W^+ W^-)=122$ fb, $\sigma(\ell^- \nu W^- W^+)=58$ fb, $\sigma(\ell^+ \ell^- W^+ Z)=9$ fb, and $\sigma(\ell^- \ell^+ W^- Z)=4$ fb. 
Without  event selection criteria and event kinematic cuts, we naively  show the expected number of events with 100 fb$^{-1}$  in Table~\ref{tab:event}. In this work, we just show the potential for discovering the VLLs, and the detailed event simulation with kinematic cuts will be given elsewhere. 
\begin{table}[b]
\begin{center}
\caption{ Number of events for the processes  in Eqs.~(\ref{eq:pair}) and (\ref{eq:single}), where  a luminosity of  100 fb$^{-1}$  and the center-of-mass energy  of $13$ TeV are used.}
\label{tab:event}
\begin{tabular}{cccccc} \hline \hline
Final state & $ \ell^- W^- \ell^+ W^+$ \quad & $ \ell^+ \nu W^+ W^-$ \quad & $ \ell^- \nu W^- W^+$ \quad & $\ell^+ \ell^- W^+ Z$ \quad & $\ell^- \ell^+ W^- Z$  \\
$m_{\Psi} = 300$ GeV & 7600 & 12200 & 5800 & 900 & 400   \\  
$400$ GeV & 2400 & 4000 & 1760 & 297 & 121  \\
$500$ GeV & 928 & 1620 & 648 & 119 & 44.7   \\  
$600$ GeV & 405 & 727 & 273 & 53.6 & 18.9   \\  
$700$ GeV & 192 & 355 & 126 & 26.2 & 8.67   \\  
$800$ GeV & 97.2 & 183 & 61.8 & 13.5 & 4.26   
 \\ \hline \hline
\end{tabular}
\end{center}
\end{table}
 
We also examine  $\Psi^{\pm \pm}$ production via flavor-changing interactions: $p p \to \Psi_1^{\pm \pm} \ell^\mp$. 
 Although the cross section is $O(1)-O(10)$ fb for $m_\Psi = 300-400$ GeV, as shown in Fig.~\ref{fig:sigma}, it would be possible to find the signal with a sufficiently large luminosity.
 This process will thus also be important,  since we could test the flavor-changing coupling $v Y/m_\Psi$ in collider experiments. However, detailed analysis of this  is left for  future work.

In summary, we investigated Higgs-portal lepton flavor violation by introducing two triplet vector-like leptons to the SM;  one is the hypercharge $Y=-1$  and the other is $Y=0$. The model has the Higgs-mediated and $Z$-mediated flavor-changing neutral currents at the tree level. When the bounds from rare $Z\to \bar \ell_i \ell_j$  decays are smeared out, the invisible $Z$ decays become the dominant constraints. As a result, the branching ratio for $h\to \mu \tau$ can be up to $10^{-4}$, muon $g-2$ is two orders of magnitude smaller than the current data, and $BR(\tau\to \mu \gamma)$ is of $O(10^{-12})$. The deviation of signal strength from the SM prediction in $\tau\tau$ mode  is $12\%$. We analyze the production channels for the predicted doubly charged lepton. We find that the interesting production channels in $pp$ collisions are $pp\to (\Psi^{--}_{1} \Psi^{++}_1, \Psi^{\pm\pm}_1 \Psi^\mp_1)$. Both  single and pair production cross sections of $\Psi^{++}_1$ are comparable and can be a few hundred fb. The main decay channel for the doubly charged lepton is $\Psi^{\pm\pm} \to \ell^\pm W^\pm$, while the heavy singly charged lepton is $\Psi^\pm_1 \to \nu W^\pm, \ell^\pm Z$.  We then summarize possible signatures of our model with the expected number of events for 100 fb$^{-1}$ luminosity at the 13 TeV LHC. \\

\noindent{\bf Acknowledgments}

The work of C. H. C. was supported by the Ministry of Science and Technology of Taiwan, Republic of China, under Grant No. MOST-103-2112-M-006-004-MY3.

\end{document}